\documentclass{PoS}
\usepackage{xspace}
\usepackage{cite,mcite}
\usepackage{wrapfig}

\title{Hadron Production at Fixed Target Energies and Extensive Air Showers}

\ShortTitle{Hadron Production at Fixed Target Energies and Extensive Air Showers}

\author{\speaker{M.\ Unger} for the NA61/SHINE Collaboration\\
        Karlsruher Institut f\"ur Technologie\\
        E-mail: \email{Michael.Unger@kit.edu}}

\abstract{NA61/SHINE is a fixed-target experiment to study hadron
  production in hadron-nucleus and nucleus-nucleus collisions at the
  CERN SPS. Due to the very good acceptance and particle
  identification in forward direction, NA61/SHINE is well suited for
  measuring particle production to improve the reliability of air
  shower simulations. Data with proton and pion beams have been taken
  in 2007 and 2009. First analysis results for the pion yield in
  proton-carbon interactions at 31 GeV will be shown and compared to
  predictions from models used in air shower simulations.}

\FullConference{35th International Conference of High Energy Physics - ICHEP2010,\\
		July 22-28, 2010\\
		Paris France}

\newcommand{\pip}{\ensuremath{\pi^+}}
\newcommand{\pim}{\ensuremath{\pi^-}}

\newcommand{\FlukaLong}{{\scshape Fluka2008}\xspace}
\newcommand{\Fluka}{{\scshape Fluka}\xspace}
\newcommand{\UrqmdLong}{{\scshape Urqmd1.3.1}\xspace}
\newcommand{\Urqmd}{{\scshape Urqmd}\xspace}
\newcommand{\GheishaLong}{{\scshape Gheisha2002}\xspace}

\newcommand{\Gheisha}{{\scshape Gheisha}\xspace}
\newcommand{\Corsika}{{\scshape Corsika}\xspace}

\begin{document}

\section{Introduction}
Ultra-high energy cosmic rays initiate extensive air showers (EAS)
when they collide with the nuclei of the atmosphere.  The
interpretation of EAS data as for instance recorded by the Pierre
Auger Observatory~\cite{Auger} or the KASCADE air shower
array~\cite{KASCADE} relies to a large extent on the under-
\begin{wrapfigure}[17]{r}[0pt]{6cm}
\includegraphics[width=\linewidth]{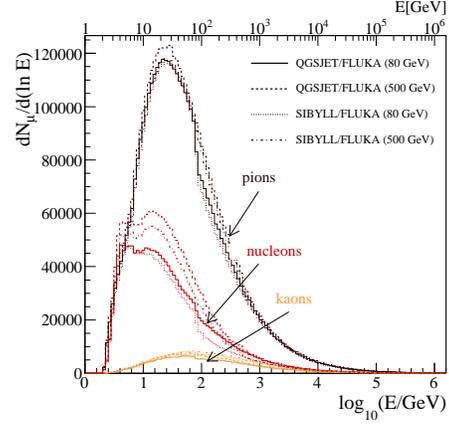}
\caption[particle energies]{
Particle types and energies involved in the last interaction
leading to muon detected at ground level ($E_0=10^{19}$~eV, detection
distance $\approx$~1~km)~\cite{Maris:2009x1}.}
\label{grandmothers}
\end{wrapfigure}
standing of these air showers and specifically on the correct modeling
of hadron-air interactions that occur during the shower
development. The relevant particle energies span a wide range from
primary energies of $\gtrsim 10^{20}$~eV down to energies
of~$10^{9}$~eV. The mesons that decay to muons at ground level
typically originate from low energy interactions in the late stages of
an air shower. Depending on the primary energy and detection distance,
the corresponding interaction energies are between 10 and 1000~GeV
(cf.~Fig~\ref{grandmothers}). As it has been noted in
e.g.\ \cite{Heck:2003br, Drescher:2003gh, Meurer:2005dt,
  Maris:2009uc}, the modeling of low energy interactions contribute
at least 10\% to the overall uncertainty of the predicted muon number at ground.
Since these energies are within the reach of fixed target experiments,
precise measurements of hadronic particle production at, for instance,
the Super Proton Synchrotron at CERN can help to diminish the
uncertainties of air shower simulations.

An example of current difficulties to describe air shower measurements
at ultra-high energies is the excess of the number of ground level
muons wrt.\ to air shower
simulations~\cite{qgsjet01,qgsjetII,sibyll2.1} as reported
by the Pierre Auger Observatory~\cite{augerMuon}.
\begin{figure}[b!]
\includegraphics[width=.545\linewidth]{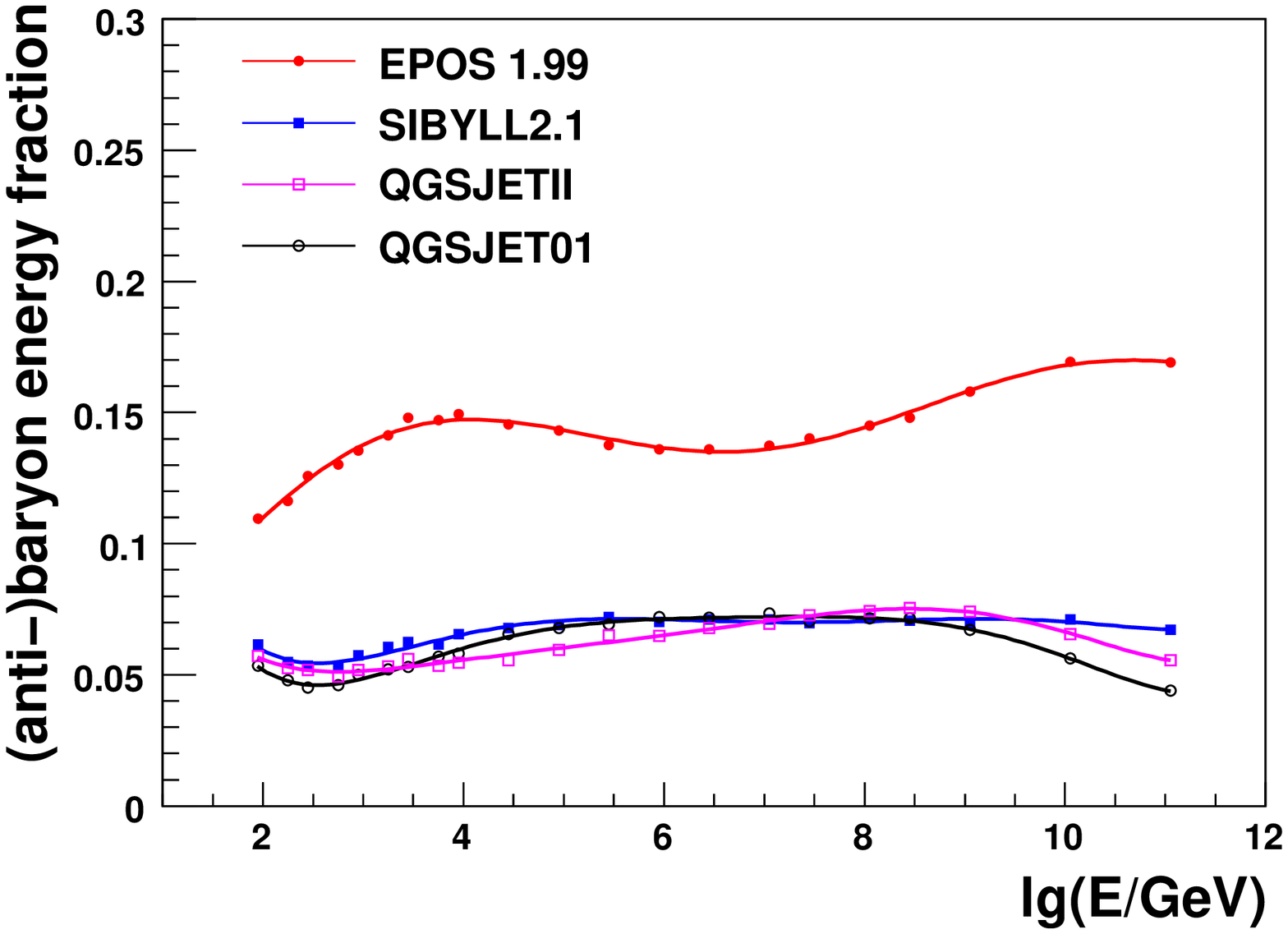}
\includegraphics[width=.455\linewidth]{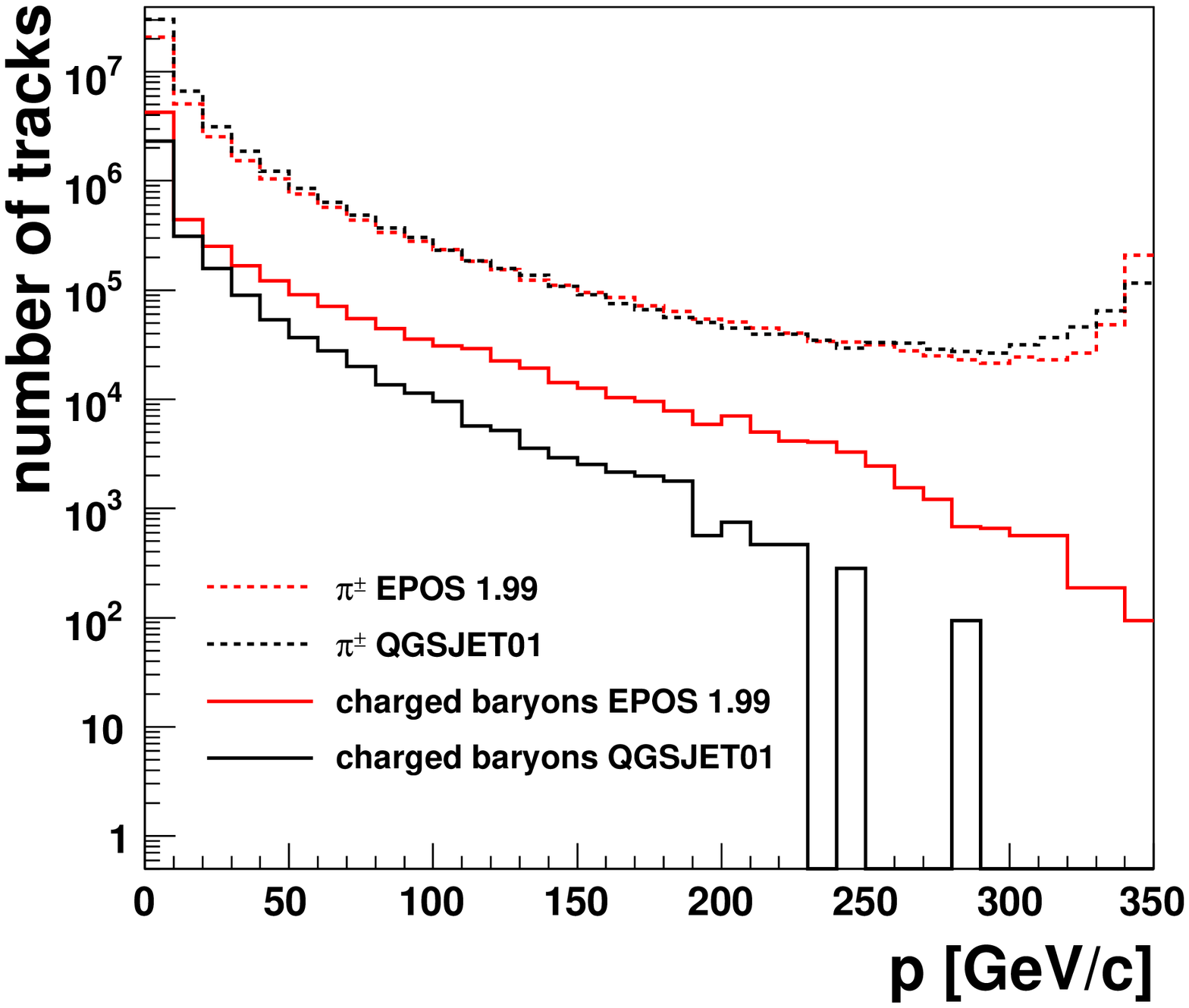}
\caption{
Left panel: Energy fraction of produced baryons and anti-baryons
in $\pi$-air collision as a function of pion momentum. Right panel:
Expected number of charged tracks as a function of secondary momentum
for the NA61 $\pi^-$-C data set at 350 GeV/c.
}
\label{baryonPrediction}
\end{figure}
A solution to this inconsistency was proposed in~\cite{epos}, where it
was pointed out that an increased production of baryons and anti-baryons
in hadron-air collisions, would lead to an increase in the number of muons
at ground level. As can be seen in Fig.~\ref{baryonPrediction},
an enhanced (anti-)baryon production, as e.g.\ currently implemented in
the {\scshape Epos}-model, should be easily distinguishable from previous
model assumptions at fixed target energies.\\

Unfortunately, there exist no comprehensive and precise
particle production measurements for the most numerous projectile
in air showers, the $\pi$-meson. Therefore, new data with pion
beams at 158 and 350 GeV/c on a thin carbon target (as a proxy for nitrogen)
has been recently collected by the NA61 experiment at the SPS.

\section{\label{sec:shine}The NA61/SHINE Experiment}

  NA61/SHINE (SHINE = SPS Heavy Ion and Neutrino
  Experiment)~\cite{na61} is an experiment at the CERN SPS using the
  upgraded NA49 hadron spectrometer. Among its physics goals are
\begin{wraptable}[18]{r}[0pt]{5.4cm}
        \begin{tabular}{cccl}\hline\hline
           & p & year & $N_\mathrm{int}$ \\
           &  [GeV/c] &  & $[10^6]$ \\\hline
    $\pi^-$-C    &   158     & 2009 & \phantom{0}3.6 \\
    $\pi^-$-C    &   350     & 2009 & \phantom{0}4.7 \\
          p-C    &   31      & 2007 & \phantom{0}0.6 \\
          p-C    &   31      & 2009 & \phantom{0}4.8 \\
          p-p    &    13     & 2010 & \phantom{0}0.6 \\
          p-p    &    20     & 2009 & \phantom{0}1.7 \\
          p-p    &    30     & 2009 & \phantom{0}2.6 \\
          p-p    &    40     & 2009 & \phantom{0}4.2 \\
          p-p    &   80     & 2009 & \phantom{0}3.6 \\
          p-p    &   158     & 2009 & \phantom{0}2.8 \\
          p-p    &   158    & 2010 & 43.9 \\\hline\hline
        \end{tabular}
\label{tab:data}
\caption{Number of thin target interaction triggers collected
         by NA61.}
\end{wraptable}
  precise hadron production measurements for improving calculations of
  the neutrino beam flux in the T2K neutrino oscillation
  experiment~\cite{T2K} as well as for more reliable simulations of
  cosmic-ray air showers. Moreover, p+p, p+Pb and nucleus+nucleus
  collisions will be studied extensively to allow for a study of
  properties of the onset of de-confinement and search for the
  critical point of strongly interacting matter.

  The NA61 detector uses large time-projection-chambers
  to measure the particle charges and momenta as well as their
  energy loss for particle type identification.
  Large scintillator walls provide an estimate of the particle's
  squared mass from the time-of-flight through the detector.
  The momentum resolution, $\sigma(1/p)=\sigma(p)/p^2$,
  is about $10^{-4}$~(GeV/c)$^{-1}$ at full magnetic field and the
  tracking efficiency is better than 95\%.

 NA61 started data taking in 2007.
 After a pilot run with proton on
 carbon at 31 GeV/c, the data acquisition system has been
  upgraded during 2008 to increase the event processing rate
 by a factor of $\approx 10$.
 In the last two years, NA61 took data at beam energies from 13 to
 350 GeV with proton and pion projectiles and proton and carbon targets
 (cf.\ Tab.~\ref{tab:data}).

\section{\label{sec:method}Data Analysis}
In this paper we present preliminary results on the inclusive
production of positive and negative pions from p+C
interactions at 31~GeV/c recorded during the 2007 pilot run.
The pion spectra have been obtained using three
independent analysis techniques: Firstly, with the so-called $h^{-}$
method all negative hadrons produced in a collision are assumed to be
pions and the small contribution of other species is corrected for using
simulations. Due to the large contribution from protons, this method
can only be applied to determine the $\pim$ spectra.  Secondly, with the
$\mathrm{d}E/\mathrm{d}x$ method $\pi$-mesons are identified
explicitly using the energy deposit measured in the TPCs. This method
works only in momentum regions where Bethe-Bloch bands do not overlap.
Finally, using the $\mathrm{d}E/\mathrm{d}x$-plus-ToF
method, $\pi^{\pm}$ can be identified over a wide momentum range using
the combination of energy loss and $m^2$ from
the time-of-flight measurement.
This method provides the most precise particle identification, but it
is limited to the angular acceptance of the time-of-flight detectors.
Using a combination of all three methods, NA61 is able to measure $\pi^{\pm}$-spectra
with a large acceptance in angle and momentum.

\section{\label{sec:results}Results}
\begin{figure}[t!]
\includegraphics[width=.245\linewidth]{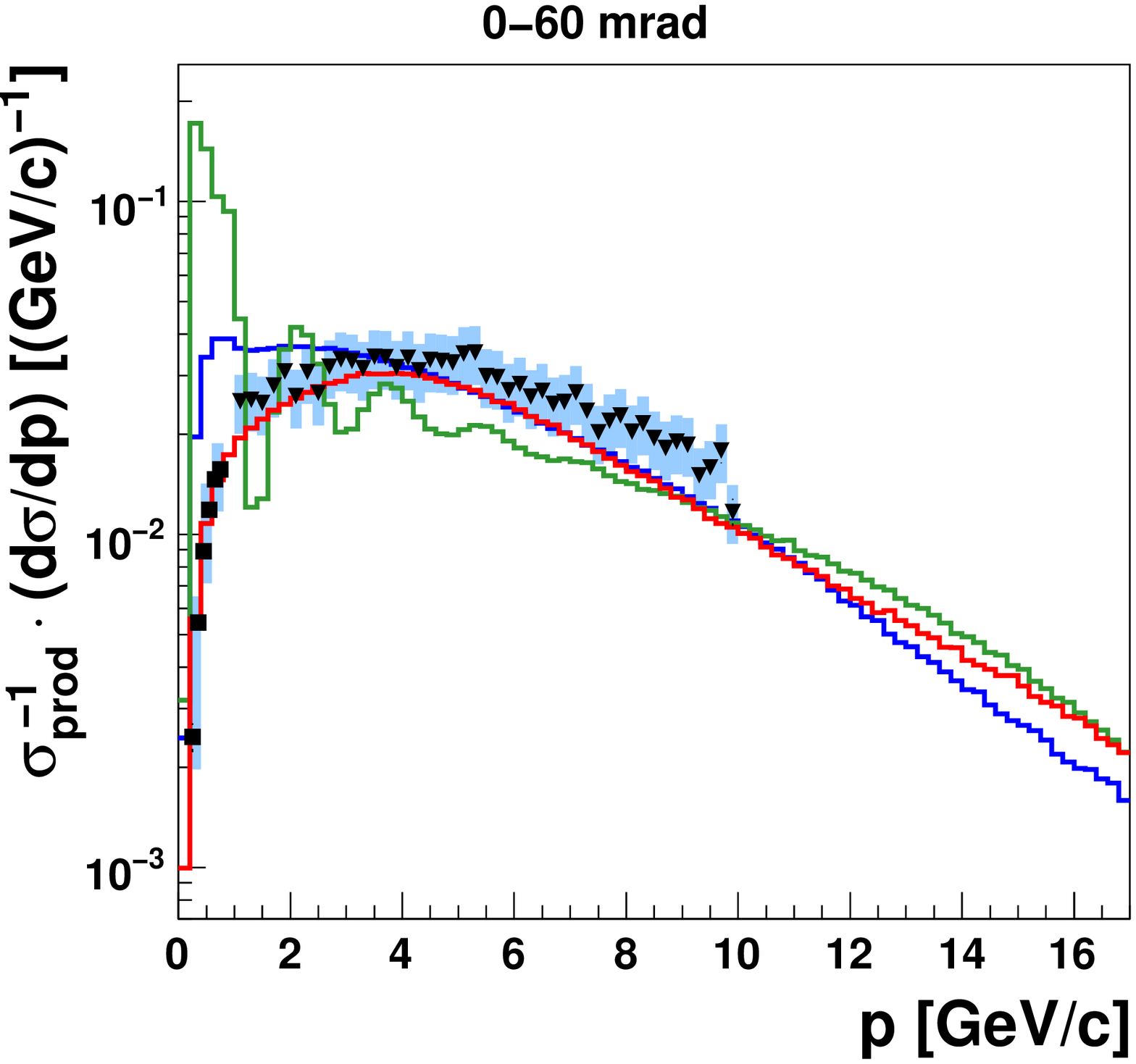}
\includegraphics[width=.245\linewidth]{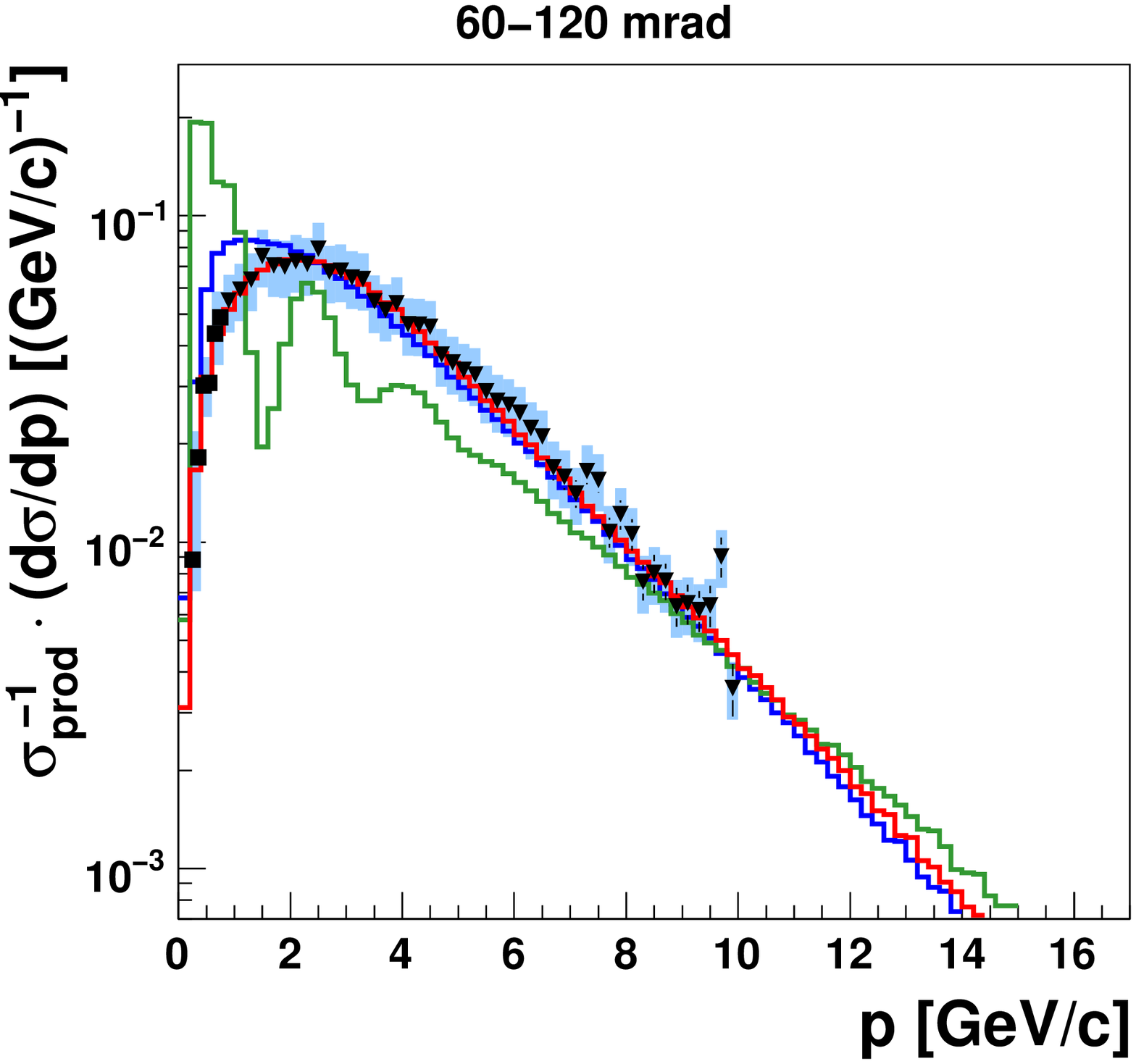}
\includegraphics[width=.245\linewidth]{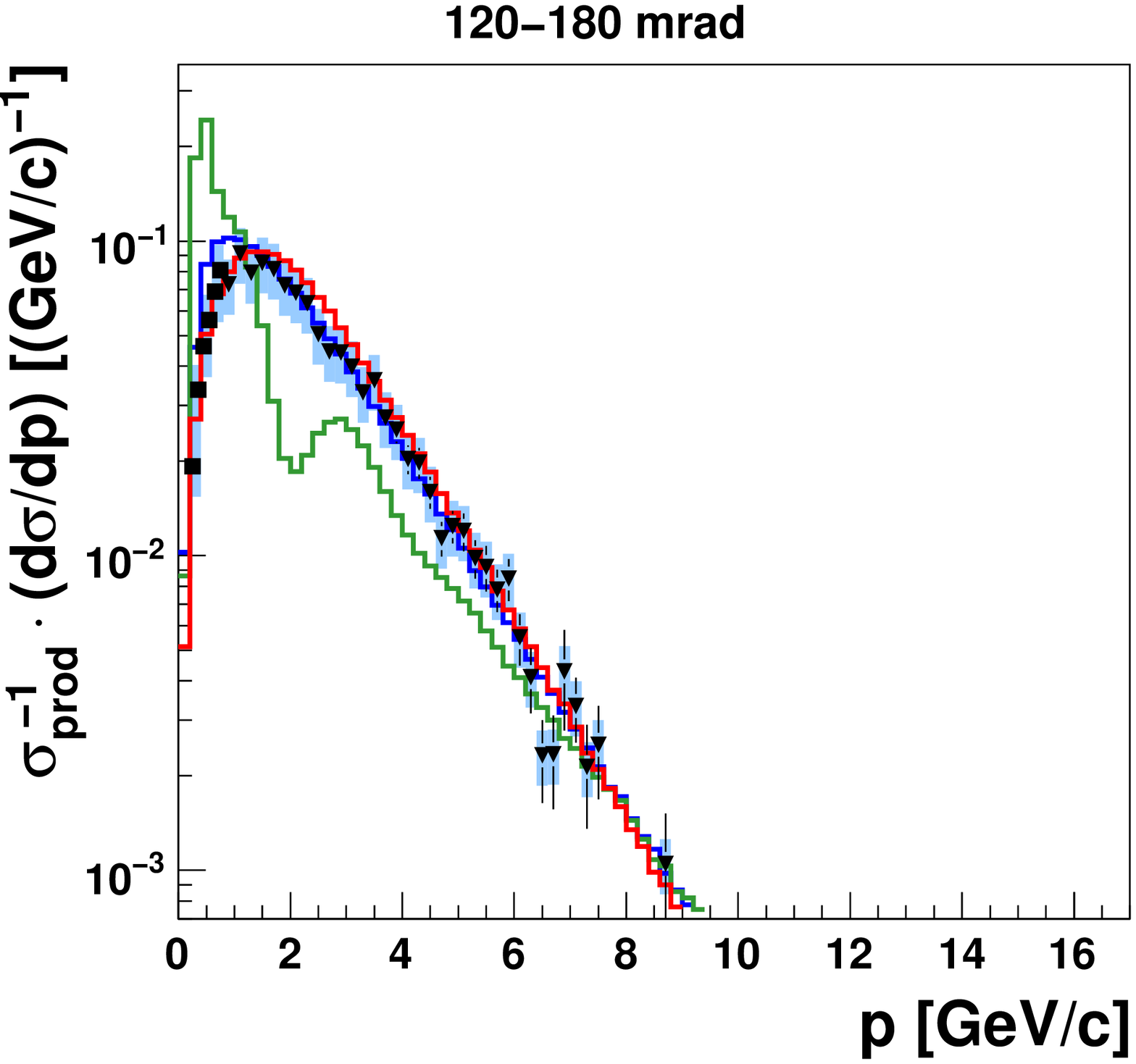}
\includegraphics[width=.245\linewidth]{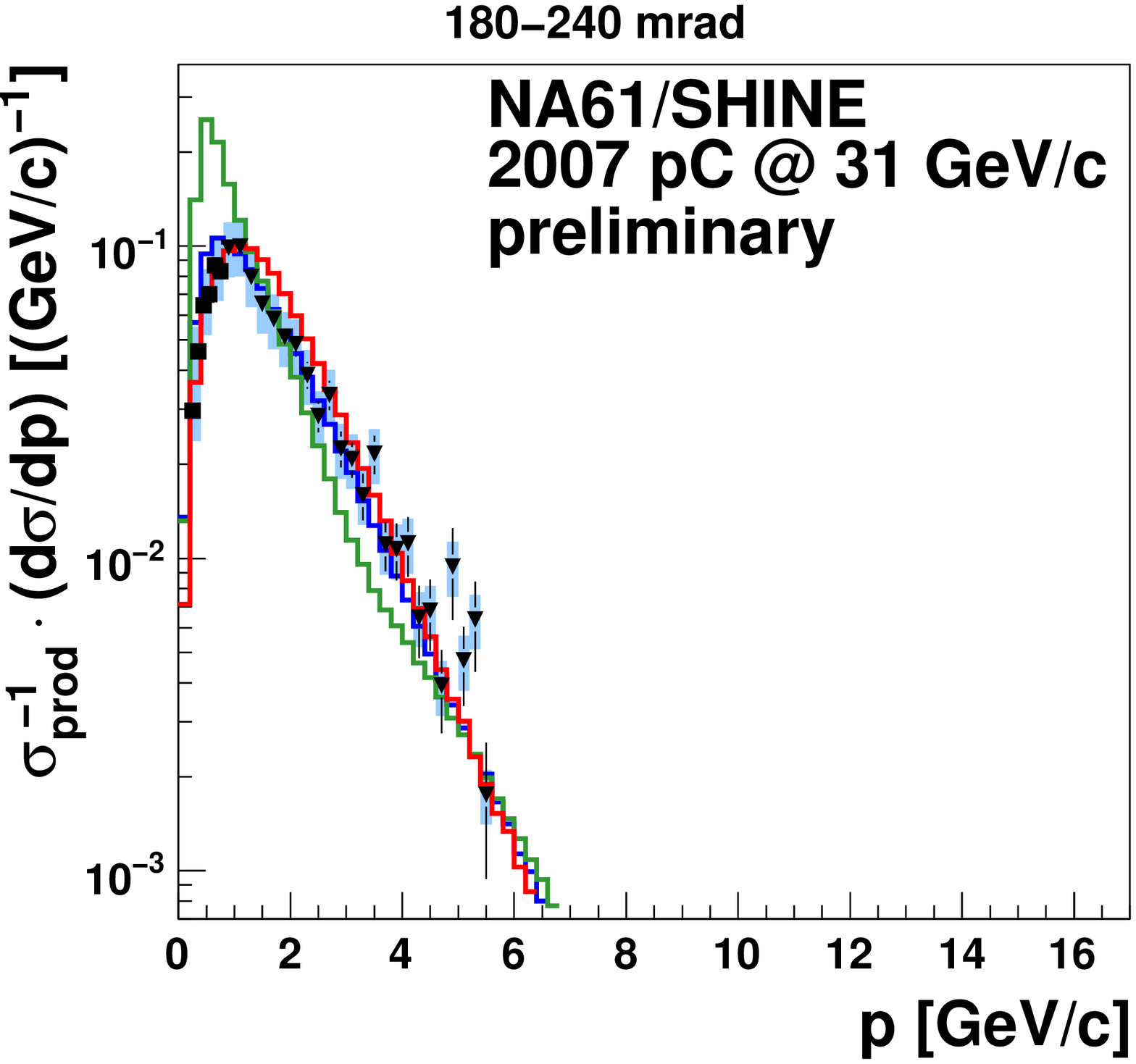}\\
\includegraphics[width=.245\linewidth]{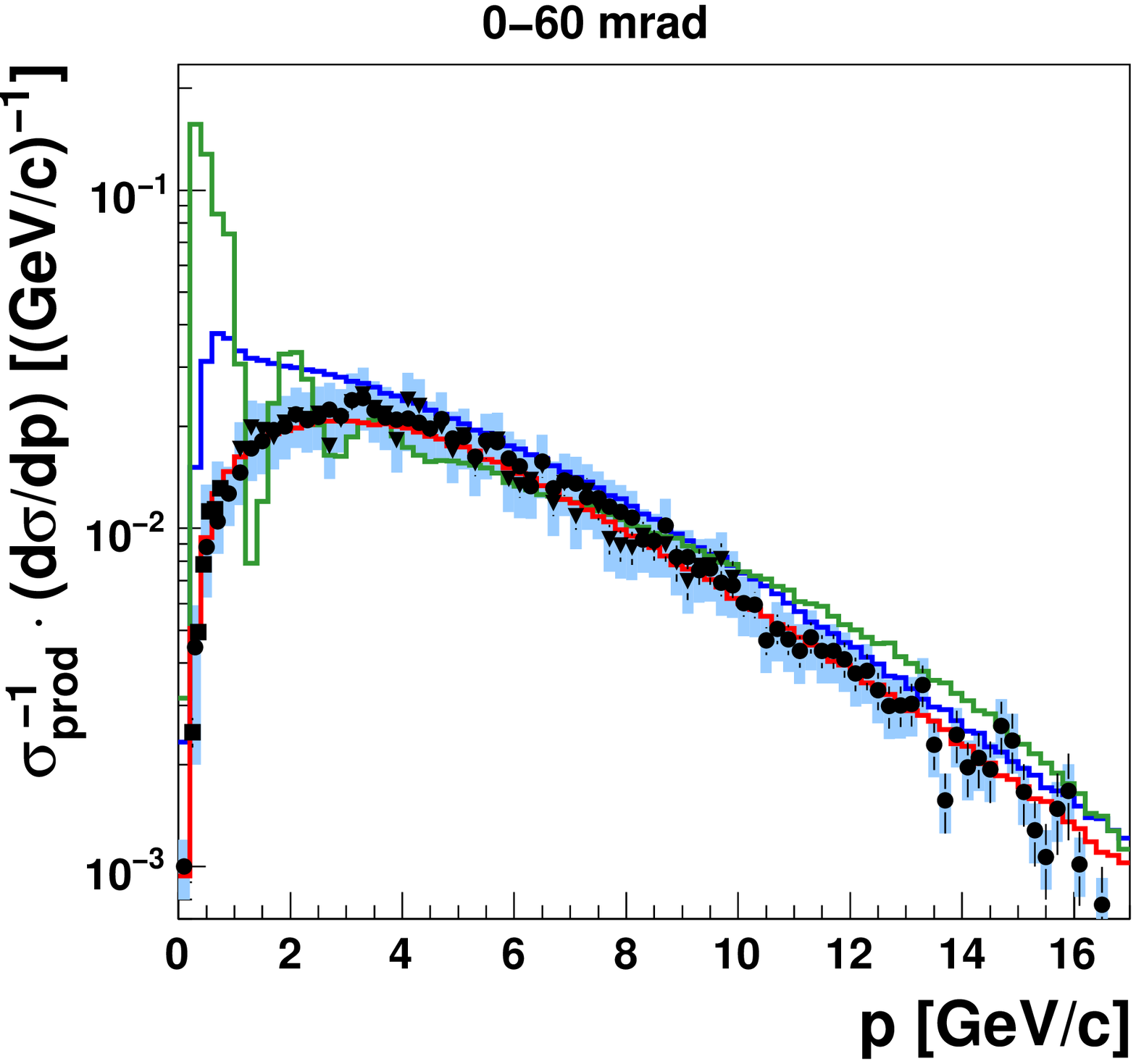}
\includegraphics[width=.245\linewidth]{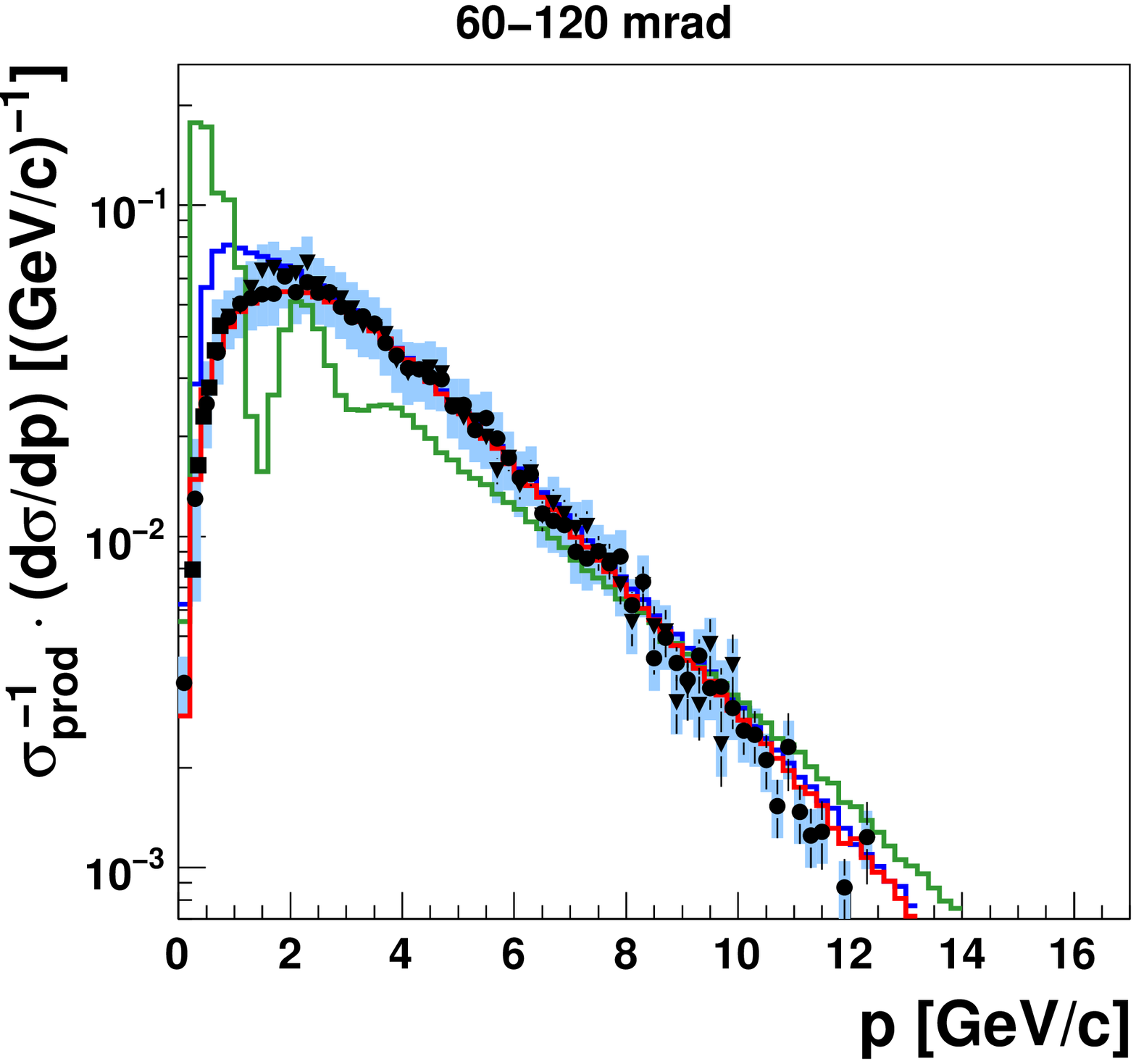}
\includegraphics[width=.245\linewidth]{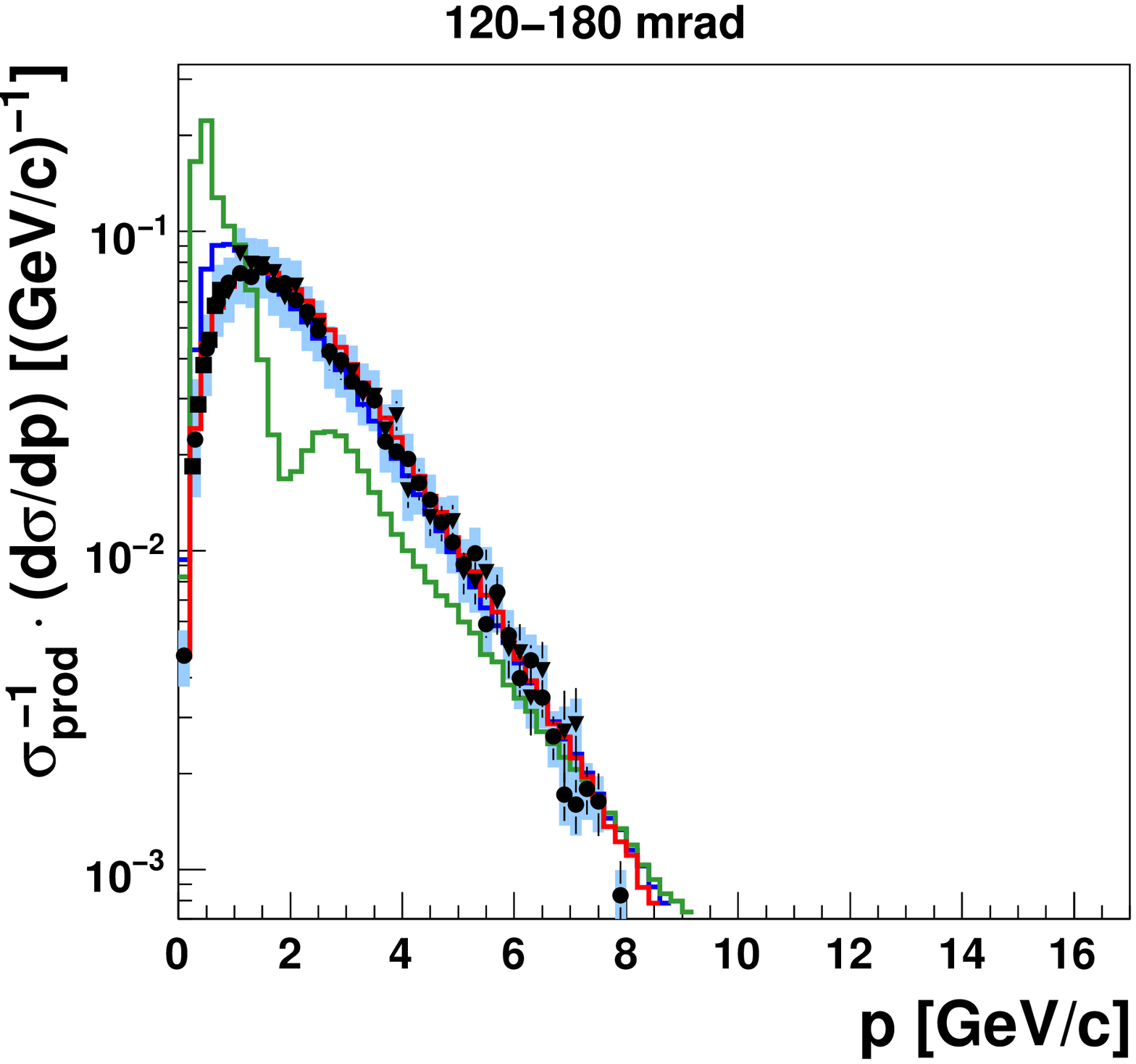}
\includegraphics[width=.245\linewidth]{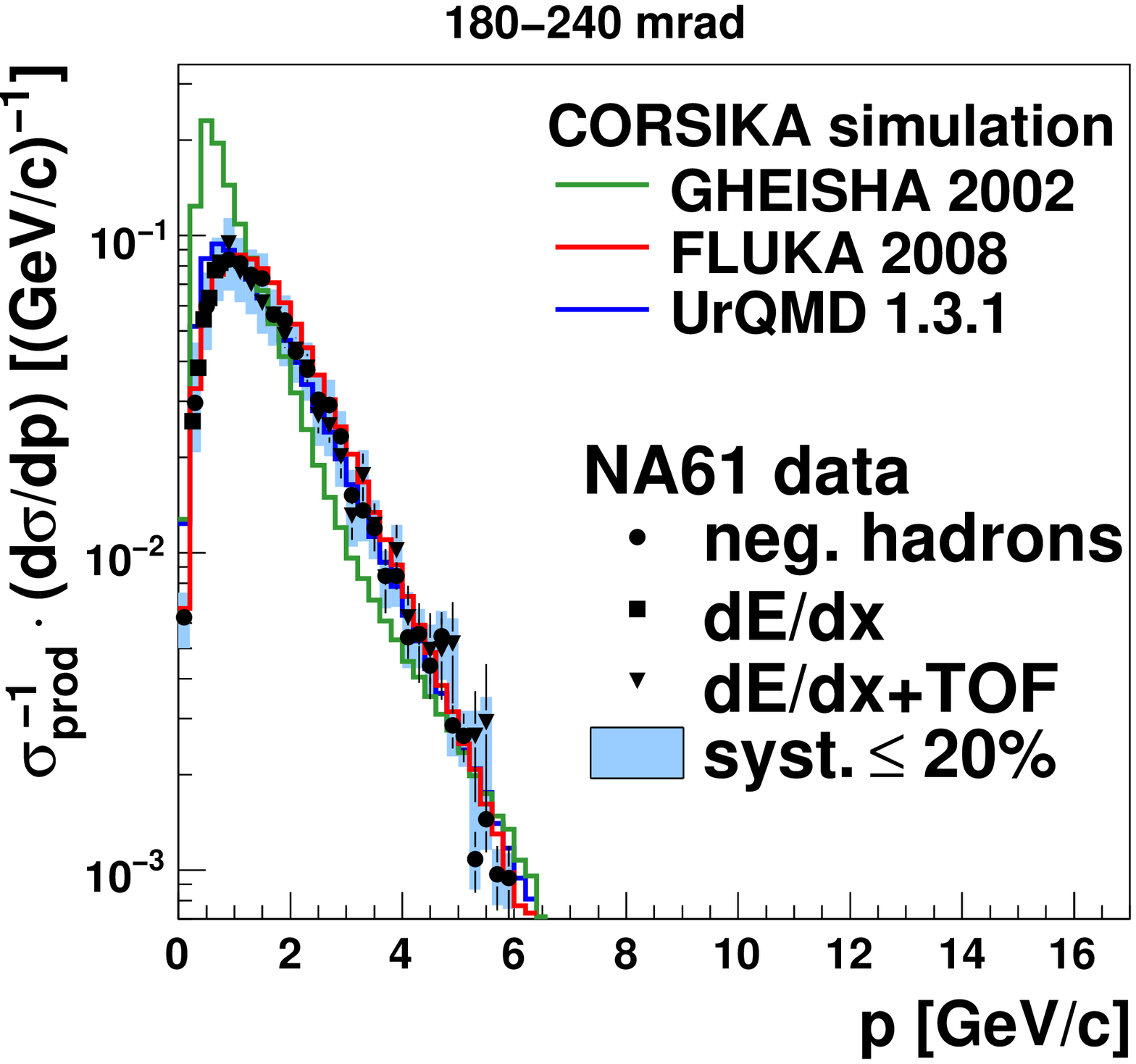}
\caption{
The $\pip$ (upper panel) and $\pim$ (lower panel)
spectra in p+C interactions at 31~GeV/c compared
to \FlukaLong (red histogram), \GheishaLong (green histogram), and \UrqmdLong (blue histogram) models.
The spectra are normalized to the total number of inelastic collisions.
}
\label{modelComp}
\end{figure}

Preliminary $\pi^{\pm}$-spectra in p+C
interactions at 31~GeV/c from the 2007 pilot run are presented
in Fig.~\ref{modelComp}. The systematic uncertainty of these preliminary
spectra is estimated to be $\le$~20\%.

As a first application of these measurement, it
is interesting to compare the spectra to the predictions of event
generators for hadronic interaction.  Here we concentrate on hadronic
interaction models that have been frequently used for the
interpretation of cosmic ray data, i.e.\ \FlukaLong~\cite{Fluka},
\UrqmdLong~\cite{Urqmd} and \GheishaLong~\cite{Gheisha}. All three
models are part of the \Corsika~\cite{Corsika} framework for the
simulation of air showers and are typically used to generate
hadron-air interactions at energies below 80~GeV.  To assure that all
relevant settings of the generators are identical to the ones used in
air shower simulations, we simulated single p+C interactions directly
with \Corsika in the so-called {\itshape interaction test} mode.  As
can be seen in Fig.~\ref{modelComp}, \Gheisha simulations can not
describe our measurements at all production angles and thus this
measurement corroborates earlier findings of the short-comings of
\Gheisha (see e.g.~\cite{Heck:2003br}).  The \Urqmd generator can
describe our data better, but fails to reproduce the spectra at low
momenta and production angles. The best agreement is found for the
spectra generated with \Fluka, that show a good overall agreement with
our data.

\section*{Acknowledgments}
This work was supported by
the Hungarian Scientific Research Fund (grants OTKA 68506 and 79840),
the Polish Ministry of Science and Higher Education (grants 667/N-CERN/2010/0,
N N202 1267 36, DWM/57/T2K/2007),
the Federal Agency of Education of the Ministry of Education and Science
of the Russian Federation (grant RNP 2.2.2.2.1547) and
the Russian Foundation for Basic Research (grants 08-02-00018 and 09-02-00664),
the Ministry of Education, Culture, Sports, Science and Technology,
Japan, Grant-in-Aid for Scientific Research (grants 18071005, 19034011,
19740162),
the German Research Foundation (grant GA 1480/2-1),
the Institut National de Physique Nucl\'eaire et Physique des Particules
(IN2P3, France),
Swiss Nationalfonds Foundation (grant 200020-117913/1)
and ETH Research Grant TH-01 07-3.

\end{document}